\documentclass[12pt,titlepage]{article}
\usepackage{amssymb,epsf}
\makeatletter
\@addtoreset{equation}{section}
\makeatother
\renewcommand{\a}{\alpha}
\renewcommand{\b}{\beta}
\renewcommand{\d}{\delta}
\newcommand{\D}{\Delta}
\newcommand{\e}{\varepsilon}
\newcommand{\eps}{\varepsilon}
\renewcommand{\l}{\lambda}        
\renewcommand{\L}{\Lambda}
\newcommand{\Le}{\Lambda_1}       
\newcommand{\Lz}{\Lambda_2}
\newcommand{\f}{{\phi}}
\renewcommand{\t}{\tau}  
\renewcommand{\S}{\Sigma}      
\newcommand{\LL}{{\cal L}}
\newcommand{\FF}{{\cal F}}
\newcommand{\NN}{{\cal N}}
\newcommand{\equ}[1]{(\ref{#1})}
\newcommand{\beq}{\begin{equation}}
\newcommand{\eeq}{\end{equation}}
\newcommand{\eeqn}[1]{\label{#1}\end{equation}}
\newcommand{\bea}{\begin{eqnarray}}
\newcommand{\eea}{\end{eqnarray}}
\newcommand{\eean}[1]{\label{#1}\end{eqnarray}}
\newcommand{\lt}{\left}
\newcommand{\rt}{\right}
\newcommand{\non}{\nonumber\\}
\newcommand{\6}{\partial}
\newcommand{\2}{\frac{1}{2}}
\newcommand{\4}{\frac{1}{4}}
\newcommand{\8}{\frac{1}{8}}
\begin{document}
%
%
%
\setcounter{page}{0}
\pagestyle{empty}
\vspace*{15mm}
\begin{flushright}{CERN-TH/99-232\\ hep--th/9908143}
\end{flushright}
\vspace*{5mm}
\begin{center}
{\bf The Prepotential of \\
$\bf \NN=2$ $\bf SU(2)\times SU(2)$ Supersymmetric Yang--Mills Theory\\ 
with Bifundamental Matter}\\
\vspace*{1cm} 
Ulrike Feichtinger$^{*}$ \\
\vspace{0.3cm}
Theory Division, CERN,\\ 
Geneva, Switzerland\\
\vspace*{2cm}  
{\bf Abstract} \\ \end{center}
\vspace*{5mm}
\noindent
%
We study the non-perturbative, instanton-corrected effective action of the 
$\NN=2$ $SU(2)\times SU(2)$ supersymmetric Yang--Mills theory with a 
massless hypermultiplet in the bifundamental representation. 
Starting from the appropriate hyperelliptic curve, we determine the periods 
and the exact \mbox{holomorphic} prepotential in a certain
weak coupling expansion. We discuss the dependence of the solution on the 
parameter $q=\frac{{\L_2}^2}{{\L_1}^2}$ and several other interesting 
properties.

\vspace*{.3cm} 
\noindent 
\rule[.1in]{6.5cm}{.002in}\\
\noindent
$^{*}$ Ulrike.Feichtinger@cern.ch
\vspace*{0.3cm}
\begin{flushleft} CERN-TH/99-232\\
August 1999
\end{flushleft}
\vfill
\eject
\pagestyle{plain}
%
%
%
\section{Introduction} \label{intro}
%
%
%

Many field theoretical results have been obtained by considering brane 
configurations in string theory and M-theory. In particular M-theory 
five-branes provide a very fruitful approach to $\NN=2$~\cite{KLMVW,witten}
and $\NN=1$~\cite{wittenii,HOO,BIKSY} \mbox{supersymmetric} field theories. 
In this note we study the case of product gauge group $SU(2) \times SU(2)$ 
with a hypermultiplet in the bifundamental \mbox{representation}. 
Seiberg--Witten curves of theories with product gauge groups have been 
obtained from the M-theory~\cite{witten} and the geometric engineering 
\mbox{approach}~\cite{KMV}. The precise \mbox{relation} between the moduli of 
the curve and the moduli of the field theory has been described 
in~\cite{randall}.

Yet another account of supersymmetric field theories with product 
gauge groups of the form $G=\prod_i SU(N_i)$ has been obtained 
in the framework of Calogero--Moser systems, using certain limits
of $SU(N)$ \mbox{supersymmetric} Yang--Mills theory with matter in the adjoint 
representation~\cite{dhoker}. In this \mbox{approach} the dynamical scales of 
the different gauge group factors are all equal to the scale of the 
underlying $SU(N)$ group. One-instanton \mbox{predictions} for the prepotential
of $\NN=2$ supersymmetric field theories with gauge group 
$SU(N_1) \times SU(N_2)$ and massless matter in the bifundamental 
representation have been made by using a perturbation expansion of the 
non-hyperelliptic curve around its hyperelliptic 
approximation~\cite{schnitzer}.

In the present paper we study the instanton expansion for the case of 
a massless hypermultiplet in the bifundamental representation of 
$SU(2) \times SU(2)$. Specifically we will compute the periods as solutions of 
the Picard--Fuchs equations and obtain in this way the holomorphic prepotential
$\FF$ that \mbox{governs} the low-energy effective action of the theory. The 
solution reproduces the \mbox{existing} results in the appropriate limits and  
has interesting properties in the general region of moduli space.

%
%
\section{The Setup}\label{setup}
%
%
%

By considering the appropriate brane configuration in M-theory~\cite{witten}
one finds that the defining polynomial for the Seiberg--Witten curve of $\NN=2$
supersymmetric Yang--Mills theory with a gauge group of the form 
$\prod_i^n SU(k_i)$ is given by
\beq
P(x,t)=t^{n+1} + p_{k_1}(x)\, t^n + p_{k_2}(x)\, t^{n-1} + \cdots +
 p_{k_n}(x)\, t + c = 0\,.
\eeqn{general}
The $p_{k_i}(x)$ are polynomials of order $k_i$ in $x$, and $c$ is a constant
that \mbox{depends} on the dynamical scales $\L_i$ of the different gauge group
factors. In this \mbox{expression} the variables $x$ and $t$ correspond to the 
combinations $x^4 + i x^5$ and $e^{-(x^6 + i x^{10})/R}$ in the notation 
of~\cite{witten}.

For the case of $SU(2) \times SU(2)$ with a massless hypermultiplet in the 
bifundamental representation, the explicit expressions for the polynomials 
$p_{k_i}(x)$ in~\equ{general} have been derived by using symmetries and 
classical limits~\cite{randall}. The curve for the massive case is given by
\beq
P(x,t)={\Le}^2 t^3+t^2\lt[x^2 -u +\frac{{\Le}^2}{2}\rt]+t\lt[(x+m)^2 -v+
\frac{{\Lz}^2}{2}\rt]+{\Lz}^2 = 0.
\eeqn{curve}
Here $u$ and $v$ are the moduli of the two gauge group factors, $\L_i$ are the 
\mbox{corresponding} dynamical scales, and $m$ is the bare mass of the 
\mbox{hypermultiplet}.
Note that the dimensionless parameter $q=\frac{{\Lz}^2}{{\Le}^2}$ cannot be 
eliminated from the curve by rescaling of the moduli, unless $q=1$.

\begin{figure}[h]
\begin{center}
\unitlength1cm
\begin{picture}(6.5,2.5)
\put(0,0.6){\line(1,0){2}}
\put(0,1.1){\line(1,0){2}}
\put(2,1.5){\line(1,0){2}}
\put(2,0.8){\line(1,0){2}}
\thicklines
\put(0,0){\line(0,1){2}}
\put(2,0){\line(0,1){2}}
\put(4,0){\line(0,1){2}}
\thinlines
\put(5,.5){\vector(1,0){1}}
\put(5,.5){\vector(0,1){1}}
\put(4.9,1.7){$x$}
\put(6.2,.4){$x^6$}
\end{picture}
\caption{The brane configuration that gives rise to a four-dimensional field 
theory with gauge group $SU(2) \times SU(2)$ and a hypermultiplet in the 
\mbox{bifundamental} representation. Vertical lines represent five-branes, 
horizontal lines are four-branes.}
\label{su2xsu2} 
\end{center}
\end{figure}
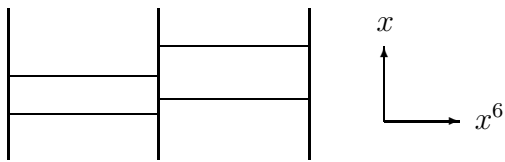
Exchanging the two gauge group factors, 
i.e.~$u \leftrightarrow v$, ${\Le} \leftrightarrow {\Lz}$, 
$t \leftrightarrow \frac{1}{t}$, and $x \leftrightarrow -(x+m)$, leaves the 
curve~\equ{curve} invariant. Pulling the rightmost five-brane to $x^6=\infty$ 
gives the brane configuration of a theory with gauge group $SU(2)$ and two 
fundamental flavours with masses $m_1$ and $m_2$. To make this transformation 
manifest in the curve one has to perform the limit 
${\Lz} \to 0$, $v \to \4 (m_1-m_2)^2$ and $m \to \2 (m_1+m_2)$. The theory 
with pure gauge group $SU(2)$ can be obtained by moving the two four-branes of 
one gauge group factor to $x=\infty$, so that the resulting configuration 
consists only of two four-branes stretched between two five-branes. To obtain 
the curve of this theory we have to take the limit $v \to \infty$,
$m \to \infty$ while keeping ${\Le}^2(m^2-v) = \L_0^4$ fixed.

In the case of gauge group $SU(2) \times SU(2)$ the polynomials $p_1(x)$ and
$p_2(x)$ are quadratic in $x$; we can therefore rewrite~\equ{curve} in 
hyperelliptic form by redefining%
\footnote{In the following we set the bare mass $m$ of the hypermultiplet to 
zero.} 
${x \to {{\frac{i\,y}{{\sqrt{2}}\,t\,\left(1+t\right)}}}}$:
\beq
y^2=t\lt(1+t\rt)\lt(2\,{\Le}^2\,t^3+t^2\,({\Le}^2-2\,u)+t\,({\Lz}^2-2\,v)+
2\,{\Lz}^2\rt).
\eeqn{hyperell}
The discriminant $\D$ of this curve consists of two factors, $\D=\D_1\,\D_2$, 
with
\bea
\D_1&=&4\,{{\lt(u - v + \2 {\Le}^2 - \2 {\Lz}^2\rt) }^2}, \label{discr1}\\ 
\D_2&=&8\,{\Le}^6{\Lz}^2 + 359\,{\Le}^4{\Lz}^4 + 
    8\,{\Le}^2{\Lz}^6 - 48\,{\Le}^4{\Lz}^2\,u + 148\,{\Le}^2{\Lz}^4\,u\,+\non
&& 96\,{\Le}^2\,{\Lz}^2\,{u^2} - 4\,{\Lz}^4\,{u^2} - 64\,{\Lz}^2\,{u^3} + 
    148\,{\Le}^4\,{\Lz}^2\,v - 48\,{\Le}^2\,{\Lz}^4\,v - \non
&& 304\,{\Le}^2\,{\Lz}^2\,u\,v +16\,{\Lz}^2\,{u^2}\,v - 
    4\,{\Le}^4\,{v^2} + 96\,{\Le}^2\,{\Lz}^2\,{v^2} +\non
&&16\,{\Le}^2\,u\,{v^2} - 16\,{u^2}\,{v^2} - 64\,{\Le}^2\,{v^3}.
\eea
As expected, $\D_1$ and $\D_2$ are symmetric under the exchange of the two 
gauge group factors. In general the vanishing of discriminant factors 
\mbox{describes}
points in moduli space where extra massless states appear. \mbox{Following} the
analysis of~\cite{randall}, the vanishing of $\D_1$ determines the subspace of 
the \mbox{moduli} space where the Coulomb branch meets the Higgs branch and 
the gauge group \mbox{$SU(2) \times SU(2)$} is broken to the diagonal $SU(2)$.
Therefore the extra massless states that \mbox{occur} at this point are two 
components of the bifundamental hypermultiplet. The vanishing of the second 
discriminant factor describes the singular locus where other massless states 
appear, in particular monopoles.

The classical intersection point of the Coulomb branch and the Higgs branch at 
$u=v$ gets shifted by non-perturbative effects if the two scales do not have 
equal values, i.e.~if~$q \neq 1$. This dependence is inferred by the terms 
$\frac{{\L_i}^2}{2}$ in the curve~\equ{curve} and could formally be avoided by 
choosing the defining polynomial to be
\beq
P(x,t)={\Le}^2\,t^3+t^2\lt(x^2-u+{\Le}^2\rt)+t\lt(x^2-v+{\Lz}^2\rt)+{\Lz}^2=0.
\eeqn{p(x,t)}
Then the two scales no longer explicitly appear in $\D_1$.
In order to reduce~\equ{p(x,t)} to the curve for $SU(2)$ with two flavours
we then have to perform a shift in the modulus $u$ in addition to the 
above-stated transformation: $u\to{\bar u}=u+\frac{7\,{\Le}^2}{8}$. This means 
that the origins of the two moduli spaces would simply be shifted by 
$\frac{7\,{\Le}^2}{8}$, though classically the moduli still coincide.

The polynomial~\equ{hyperell} describes a Riemann surface $\S$ of genus 2. 
If we choose a symplectic homology basis of $\S$, i.e.~$\a_i$ and $\b_i$ 
($i=1,2$) with intersection pairing $\a_i \cap \b_j = \d_{ij}$, 
$\a_i \cap \a_j = \b_i \cap \b_j =0$, we can define the period integrals in 
terms of a properly chosen meromorphic one-form $\l$:
\beq
a_i = \int_{\a_i} \l, \qquad \qquad a_{Di} = \int_{\b_i} \l.
\eeqn{integrals}
The periods are functions of the moduli $u$ and $v$, $a_i(u,v)$ is 
identified with the scalar component of the $\NN=1$ chiral multiplet in the 
$i$-th gauge group factor, $a_{Di}(u,v)$ is its dual. Once the periods 
$a_i$ and $a_{Di}$ are known, the exact quantum-corrected prepotential 
$\FF(a_k)$ of the theory can be computed by integrating the relations
\beq
a_{Di}(a_k)=\frac{\6 \FF(a_k)}{\6 a_i}.
\eeqn{Fder}
The second-order derivatives of $\FF$ with respect to $a_i$ imply the 
integrability condition
\beq
\frac{\6 a_{Di}(a_k)}{\6 a_j} = \frac{\6 a_{Dj}(a_k)}{\6 a_i}.
\eeqn{integrability}
The meromorphic one-form $\l$ of the curve~\equ{curve} is given by~\cite{KLMVW}
\beq
\l \propto \frac{x\, dt}{t}\,,
\eeq
or, after the redefinition 
$x \to {{\frac{i\,y}{{\sqrt{2}}\,t\,\left( 1 + t \right) }}}$, by
\beq
\l \propto \frac{y\, dt}{t^2\, (1+t)}\,,
\eeqn{lamda}
where the proportionality factor is determined by the requirements
\beq
\frac{\6 \l}{\6 v} =\frac{dt}{y} \quad {\rm and} \quad 
\frac{\6 \l}{\6 u} =\frac{t\,dt}{y}\,.
\eeq
It is rather tedious to perform the integrals~\equ{integrals} explicitly. 
However, the \mbox{periods} as functions of the moduli $u$ and $v$ satisfy a 
system of
partial differential equations, the so-called Picard--Fuchs equations. 
This method to obtain the periods is relatively straightforward and will be the
subject of the next \mbox{sections}.

%
%
%
\section{The Picard--Fuchs Equations}\label{PF}
%
%
%

Starting with the hyperelliptic form~\equ{hyperell} of the curve%
\footnote{For convenience we replace $t$ by $x$ in the following.}
for $SU(2) \times SU(2)$, we will derive the system of partial differential 
equations for the periods $\int_{\a_i}\l$ and $\int_{\b_i}\l$.

On a Riemann surface $\S$ of genus 2 there are two holomorphic differentials 
and two meromorphic differentials with no residues (abelian differentials of 
first and second kind)~\cite{kra}. Therefore we can choose 
$\{\frac{dx}{y},\frac{x\,dx}{y},\frac{x^2\,dx}{y},\frac{x^3\,dx}{y}\}$ as 
basis of meromorphic forms on $\S$. When considering derivatives of the 
meromorphic one-form $\l$ with respect to the moduli, only four of them will 
be linearly independent up to exact forms. The linear relations between these 
derivatives define the Picard--Fuchs equations.

Derivatives of $\l$ with respect to $u$ and $v$ involve terms of
the form $\frac{\f(x)\,dx}{y^n}$ with some polynomials $\f(x)$. In order to
express $\frac{\f(x)\,dx}{y^n}$ in terms of abelian differentials of first and 
second kind, we need a method to reduce high powers of $x$ in the numerator and
high powers of $y$ in the denominator. For the latter we use the fact that the 
discriminant of a polynomial $p(x)$ of order $n$ can always be written in the 
form~\cite{KTS}
\beq
\D = a(x)\,p(x) + b(x)\,p'(x),
\eeq
where $a(x)$, resp.~$b(x)$, is a polynomial of order $(n-2)$, resp.~$(n-1)$,
in $x$. With this formula it can easily be shown that the following relation
holds up to exact forms:
\beq
\frac{\f(x)}{y^n} = \frac{1}{\D}\,\frac{a(x)\,\f(x) + 
\frac{2}{n-2}\,\frac{d}{d x}\lt(b(x)\,\f(x)\rt)}{y^{n-2}}.
\eeq
Following~\cite{KTS} we derive a formula to reduce high powers of $x$
that is valid up to addition of total derivatives:
\bea
\frac{x^n}{y}&=& 
-\,\frac{(2 n-7)}{(2 n-3)}\,\frac{{\Lz}^2}{{\Le}^2}\,\frac{x^{n-4}}{y}-
\frac{(n-3)}{(2 n-3)}\,\frac{(3\,{\Lz}^2-2\,v)}{{\Le}^2}\,\frac{x^{n-3}}{y}-
\non
&&\frac{(2 n-5)}{(2 n-3)}\,\frac{({\Le}^2+{\Lz}^2-2\,u-2\,v)}{2{\Le}^2}\,
     \frac{x^{n-2}}{y}-\non
&&\frac{(n-2)}{(2 n-3)}\,\frac{(3\,{\Le}^2-2\,u)}{{\Le}^2}\,
\frac{x^{n-1}}{y}\,.
\eea
With these formulae we can re-express the meromorphic one-form~\equ{lamda} in 
terms of abelian differentials of first and second kind
\bea
\l&=&(2\,v-{\Lz}^2)\,\frac{dx}{y}+(4\,u+2\,v-2{\Le}^2-{\Lz}^2)\,\frac{x\,dx}{y}
+\non
  &&4\,(u-2\,{\Le}^2)\,\frac{x^2\,dx}{y}-6\,{\Le}^2\,\frac{x^3\,dx}{y}\,.
\eea
When we consider derivatives of $\l$ with respect to $u$ and $v$ up to second 
order, we find that these derivatives satisfy two linear relations as only four
out of six are linearly independent. These two relations give rise to the 
following Picard--Fuchs operators:
\bea
\LL_1 &=& (5\,{{\Le}^2} + 2\,u)\,\6_u^2 + 
    2\,\left( {{\Le}^2} - {{\Lz}^2} - 2\,u + 
    2\,v \right) \6_u\6_v - \non
&&(5\,{{\Lz}^2} + 2\,v)\,\6_v^2 + \6_u - \6_v ,\\
\LL_2&=&({{\Le}^4}+14\,{{\Le}^2}\,{{\Lz}^2}-4\,{u^2} -4\,{{\Le}^2}\,v)\,\6_u^2 
- \non
&&(14\,{{\Le}^2}\,{{\Lz}^2} + {{\Lz}^4} - 4\,{{\Lz}^2}\,u - 4\,{v^2})\,
\6_v^2 - \non
&&2\,\left( 15\,{{\Le}^2}\,{{\Lz}^2} - 2\,{{\Lz}^2}\,u - 2\,{{\Le}^2}\,v - 
    4\,u\,v \right) \6_u \6_v + 1.
\eea
Note that the first (second) operator is antisymmetric (symmetric) under the 
exchange of the two gauge group factors.

%
%
%
\section{The Periods and the Prepotential}\label{periods}
%
%
%

We are interested in the prepotential of $SU(2) \times SU(2)$  supersymmetric
Yang--Mills theory in the weak coupling regime, which is parametrized by large 
values of the moduli $u$ and $v$. As  usual, if we have two variables that 
\mbox{simultaneously} become large, we have to specify an appropriate 
coordinate patch in which the limit is well-defined. By inspection of the 
first \mbox{discriminant} factor~\equ{discr1}, we choose the variables to be 
\beq
z_1=\frac{{\Le}^2}{u} \quad {\rm and}\quad 
z_2=\frac{u-v+\2({\Le}^2-{\Lz}^2)}{{\Le}^2}.
\eeqn{variables}
This choice is not symmetric under the exchange of the two
gauge group factors, but of course there exists the analogous patch in the 
moduli space where the r\^ole of the two moduli is interchanged.
Using Frobenius' method, we find two series solutions and two logarithmic 
solutions with indices $\lt(-\2,0\rt)$ and $\lt(\2,1\rt)$ in the coordinate 
patch of the moduli space parametrized by~\equ{variables}:
\bea
w_1(z_1,z_2)&=&
	z_1^{-\2}\sum_{i,j=0}^{\infty}a_{ij}\,z_1^i\,z_2^j,\label{ser1}\\
w_2(z_1,z_2)&=&
	z_1^{\2}z_2\sum_{i,j=0}^{\infty}b_{ij}\,z_1^i\,z_2^j,\label{ser2}\\
w_3(z_1,z_2)&=& w_1(z_1,z_2)\ln{(z_1)}+
	z_1^{-\2} \sum_{i,j=0}^{\infty}c_{ij}\,z_1^i\,z_2^j,\label{log1}\\
w_4(z_1,z_2)&=& w_2(z_1,z_2)\ln{(z_1^3 z_2^2)}+
	z_1^{-\2}\sum_{i,j=0}^{\infty}d_{ij}\,z_1^i\,z_2^j.\label{log2}
\eea
The first few coefficients of these series expansions are listed in the 
appendix.

In order to get the periods we have to consider linear combinations of the
solutions that match the infrared asymptotic behaviour. To analyse these 
leading terms one can either evaluate the lowest order of the 
\mbox{integrals}~\equ{integrals} explicitly or one uses semi-classical 
relations of the gauge couplings and the integrability 
condition~\equ{integrability}.

The gauge couplings $\t_{ij}$ of the theory are related to the prepotential 
$\FF$ by
\beq
\t_{ij} = \frac{\6^2 \FF(a_k)}{\6 a_i \6 a_j} = \frac{\6 a_{Di}}{\6 a_j}.
\eeqn{tauij}
The one-loop terms of the pure couplings $\t_{ii}$ consist of the logarithmic
contributions of the massless spectrum. In each gauge group factor there
is a gauge field in the adjoint representation; moreover there is a 
hypermultiplet in the bifundamental representation. Therefore the one-loop 
contributions of the pure gauge couplings are given by
\bea
\t_{11}&\sim& -4\ln{({a_1})} + \ln{({a_1}^2-{a_2}^2)},\non
\t_{22}&\sim& -4\ln{({a_2})} + \ln{({a_1}^2-{a_2}^2)}.
\eean{taus}
In order to examine the asymptotic behaviour of the periods, we integrate the
gauge couplings $\t_{ii}$ in~\equ{taus} to obtain 
$a_{Di}(a_k)=\int\t_{ii}(a_k)\,da_i$, 
introduce a new variable $\eps={a_1}-{a_2}$, and expand around the point 
$({a_1},\eps) = (\infty, 0)$. This procedure gives the leading terms of the 
periods $a_{Di}$:
\bea
a_{D1}&\sim& - 2\,{a_1}\,\ln ({a_1}) - \e\,\ln ({a_1}) + \e\,\ln (\e)+
\ldots\,,\non
a_{D2}&\sim& - 2\,{a_1}\,\ln ({a_1}) + 3\,\e\,\ln ({a_1}) - \e\,\ln (\e)+
\ldots\,.
\eean{tauexpanded}
To compare the solutions~\equ{log1} and~\equ{log2} with these 
expansions, we have to replace the moduli $u$ and $v$ by the corresponding 
classical expressions $u = {a_1}^2$ and $v={a_2}^2$. We apply the same 
procedure as above to a linear combination of the logarithmic solutions 
$A\,w_3 + B\,w_4$ and find, in leading order:
\beq
-2\,A\,{a_1}\,\ln({a_1})+\left(A-8\,B\right)\,\eps\,\ln({a_1})+
4\,B\,\eps\,\ln(\eps)+\ldots\,.
\eeq
Hence the periods $a_{Di}$ are reproduced by the following linear combinations
of the solutions
\bea
\frac{a_{D1}}{{\Le}} &=& A_1 w_1 + B_1 w_2 + w_3 + \4 w_4,\non
\frac{a_{D2}}{{\Le}} &=& A_2 w_1 + B_2 w_2 + w_3 - \4 w_4,
\eean{adis}
where the coefficients $A_1, A_2, B_1, B_2$ are still undetermined. We also 
find that the periods $a_i$ are given by
\bea
\frac{{a_1}}{{\Le}} &=& w_1 + \4 w_2, \non
\frac{{a_2}}{{\Le}} &=& w_1 - \4 w_2.
\eean{elperiods}
It can easily be checked that in the limits of $SU(2)$ $N_f=0$, 
resp.~\mbox{$N_f=2$}, stated in section~\ref{setup}, ${a_1}$ reduces to the 
corresponding periods known in the \mbox{literature}. By imposing the 
integrability condition~\equ{integrability}, one finds that \mbox{$A_2 = A_1$} 
and $B_2=-B_1$.

The prepotential can be computed by integrating the equations~\equ{Fder}. To 
this end we have to know the magnetic periods $a_{Di}$ as functions of the
electric periods $a_k$. By inverting the series expansions of $a_k$ 
in~\equ{elperiods} we get the variables $z_i$ and in consequence the magnetic 
periods as functions of $a_k$:
\bea
a_{D1}({a_1},{a_2})&=& {a_1}\,\lt(\2\,A_1+2\,B_1+\ln{(2)}\rt)+\non
&&{a_2}\,\lt(\2\,A_1-2\,B_1-\ln{(2)}\rt)-\non
&&4\,{a_1}\ln{({a_1})}+({a_1}+{a_2})\ln{({a_1}+{a_2})}+({a_1}-{a_2})\ln{({a_1}-
{a_2})}+\non
&&\2\,\frac{{\Le}^2}{{a_1}}\lt(\frac{{a_2}^2}{{a_1}^2}-
q\,\frac{{a_1}^2}{{a_2}^2}\rt)-\non
&&\frac{{\Le}^4}{32}\lt[\2\,\frac{1}{{a_1}^7}\lt({a_1}^4-12\,{a_1}^2{a_2}^2+
15\,{a_2}^4\rt)+\rt.\non
&&\quad\lt.8\,q\,\frac{1}{{a_1}^3}-
q^2\,\frac{{a_1}}{{a_2}^6}\lt(5\,{a_1}^2-3\,{a_2}^2\rt)\rt]+\ldots\label{ad1}\\
a_{D2}({a_1},{a_2})&=&{a_1}\,\lt(\2\,A_1-2\,B_1-\ln{(2)}\rt)+\non
&&{a_2}\,\lt(\2\,A_1+2\,B_1+\ln{(2)}\rt)-\non
&&4\,{a_2}\ln{({a_2})}+({a_1}+{a_2})\ln{({a_1}+{a_2})}-({a_1}-{a_2})\ln{({a_1}-
{a_2})}-\non
&&\2\,\frac{{\Le}^2}{{a_2}}\lt(\frac{{a_2}^2}{{a_1}^2}-q\,
\frac{{a_1}^2}{{a_2}^2}\rt)-\non
&&\frac{{\Le}^4}{32}\lt[\frac{{a_2}}{{a_1}^6}\lt(3\,{a_1}^2-5\,{a_2}^2\rt)+
8\,q\,\frac{1}{{a_2}^3}\,+\rt.\non
&&\quad\lt.\2\,q^2\,\frac{1}{{a_2}^7}\lt(15\,{a_1}^4-
12\,{a_1}^2{a_2}^2+\,{a_2}^4\rt)\rt]+\ldots\label{ad2}
\eea
Notice that exchanging ${a_1}$ and ${a_2}$ transforms $a_{D1}$ into $a_{D2}$ 
and vice versa. Integrating the expression~\equ{ad1}, resp.~\equ{ad2}, with 
respect to ${a_1}$, resp.~${a_2}$, yields the prepotential
\beq
\FF=\FF_{class}+\FF_{1-loop}+\sum_{n=1}^{\infty}\FF_{n-inst}\,,
\eeqn{prepot}
with the following expressions for the first few terms:
\bea
\FF_{class}&=& ({a_1}^2+{a_2}^2)\lt(\2+\4\,A_1+B_1+\2\,\ln{(2)}\rt)+\non
&&\quad{a_1}{a_2}\lt(\2\,A_1-2\,B_1-\ln{(2)}\rt)\,,\label{Fclass}\\[2mm]
\FF_{1-loop}&=& -2\,{a_1}^2\,\ln{({a_1})}-2\,{a_2}^2\,\ln{({a_2})}+\non
&&\quad\2({a_1}+{a_2})^2\ln{({a_1}+{a_2})}+
\2({a_1}-{a_2})^2\ln{({a_1}-{a_2})}\,,\label{F1-loop}\\[2mm]
\FF_{1-inst}&=&\frac{{\Le}^2}{4}\lt({a_1}^2-{a_2}^2\rt)\lt(\frac{1}{{a_1}^2}-
q\,\frac{1}{{a_2}^2}\rt)\,,\label{F1-inst}\\[2mm]
\FF_{2-inst}&=&-\,\frac{{\Le}^4}{4}\,q\,\frac{1}{{a_1}^2}-
\,\frac{{\Le}^4}{128}\,\lt({a_1}^2-{a_2}^2\rt)\lt[\frac{{a_2}^2}{{a_1}^4}
\lt(\frac{5}{{a_1}^2}-\frac{1}{{a_2}^2}\rt)-\rt.\non
&&\quad\lt.16\,q\,\frac{1}{{a_1}^2\,{a_2}^2}+q^2\,\frac{{a_1}^2}{{a_2}^4}
\lt(\frac{1}{{a_1}^2}-\frac{5}{{a_2}^2}\rt)\rt]\,,\label{F2-inst}\\[2mm]
\FF_{3-inst}&=&-\,\frac{{\Le}^6}{384}\,({a_1}^2-{a_2}^2)\lt[
\frac{{a_2}^4}{{a_1}^8}\lt(\frac{9}{{a_1}^2}-\frac{5}{{a_2}^2}\rt)+
3\,q\,\frac{1}{{a_1}^4}\lt(\frac{5}{{a_1}^2}-\frac{1}{{a_2}^2}\rt)\rt.+\non
&&\quad\lt.3\,q^2\,\frac{1}{{a_2}^4}\lt(\frac{1}{{a_1}^2}-
\frac{5}{{a_2}^2}\rt)+
q^3\,\frac{{a_1}^4}{{a_2}^8}\lt(\frac{5}{{a_1}^2}-\frac{9}{{a_2}^2}\rt)\rt]
\,.\label{F3-inst}
\eea
As expected, all contributions to the prepotential are totally symmetric 
\mbox{under} the exchange of the two gauge group factors. The constants $A_1$ 
and $B_1$ are determined by the classical couplings. The one-loop 
\mbox{contribution} $\FF_{1-loop}$ is in agreement with the corresponding term 
of the prepotential \mbox{reported} in~\cite{dhoker} and~\cite{schnitzer}.

Up to an overall scaling factor of the two scales, $\FF_{1-inst}$ coincides 
with the one-instanton term in the prepotential of~\cite{schnitzer}.
If the bare mass $m$ of the hypermultiplet in the prepotential obtained 
in~\cite{dhoker} is set to zero, the one-instanton term coincides with 
$\FF_{1-inst}$ for $q=1$, but the terms proportional to $q$ in $\FF_{2-inst}$ 
are not present. This is a priori no contradiction, since sending the bare mass
to zero is a singular limit.

The instanton terms cannot be written as simple sums of contributions 
\mbox{stemming} from the two subgroups, but show a non-trivial mixing of 
${a_1}$ and ${a_2}$, as it is expected in the presence of a hypermultiplet in 
the bifundamental representation. In particular the $n$-th instanton term is 
not just a sum of \mbox{contributions} proportional to ${\L_i}^{2\,n}$, but all
possible combinations ${\Le}^{2\,i}\,{\Lz}^{2\,j}$ with $i+j=n$ appear in 
$\FF_{n-inst}$. Terms proportional to odd powers of the scales are forbidden 
as they would violate the anomaly-free discrete \mbox{symmetry} of the 
theory~\cite{SWII}.

It is quite striking that nearly all terms in the instanton expansion of the
\mbox{prepotential} are proportional to $\lt({a_1}^2-{a_2}^2\rt)$ and therefore
vanish for \mbox{$a_1 = \pm\,a_2$}. In either of the two cases two components 
of the hypermultiplet become massless and one expects to recover the point in 
moduli space where the Coulomb branch meets the Higgs branch and the gauge 
group is broken to the diagonal $SU(2)$. Continuity of the prepotential at the 
intersection point ensures that $\FF$ collapses to the prepotential of pure 
$SU(2)$. Indeed, if one sets $a_1=\pm\,a_2=a$ in the prepotential~\equ{prepot},
the non-vanishing terms \mbox{reproduce} the prepotential of $SU(2)$ theory 
with the dynamical scale $\L^2 = 4\,{\Le}{\Lz}$ to all calculated orders. 

In~\cite{randall} a different argumentation, namely considerations of the brane
configuration and the curve, led to the insight that the same point in 
moduli space is determined by the equation $u+\2{\Le}^2=v+\2{\Lz}^2$. At first 
sight it is quite astounding that this condition is equivalent to 
${a_1}^2={a_2}^2$, but it is clear that these two relations describe the same 
physical scenario.

\vspace{1cm}
{\Large{\bf Acknowledgements}}
\vspace{.5cm}

I would like to thank W.~Lerche and P.~Mayr, as well as P.~Kaste and 
S.~Stieberger for valuable discussions.
%
%
%
%

%
%
%
\newpage
%
%
%
\begin{appendix}
\section*{Appendix \quad Coefficients of the Solutions}\label{first}
%
%
%
%
In table~\ref{sertab} the non-vanishing coefficients of the series 
solutions~\equ{ser1} and~\equ{ser2} are listed up to fourth order in 
the variables $z_1 = \frac{{\Le}^2}{u}$ and 
$z_2 = \frac{u-v+\2({\Le}^2-{\Lz}^2)}{{\Le}^2}$. In table~\ref{logtab} the
corresponding coefficients for the logarithmic \mbox{solutions}~\equ{log1} 
and~\equ{log2} are listed.
%
%
%
{\footnotesize
\begin{table}[h]
\begin{center}
\begin{tabular}{|c|c|l|l|}
\hline
$i$ &$ j$ & $a_{ij}$ & $b_{ij}$\\[.5mm]
\hline
$0$ & $0$ & $1$ 						& $1$ \\[.5mm]
$1$ & $0$ & $\4$ 						& $-\frac{3}{4}\lt(1+\frac{2}{3}q\rt)$ \\[.5mm]
$2$ & $0$ & $-\frac{1}{32}\lt(1+8q\rt)$				& $\frac{27}{32}\lt(1+\frac{4}{3}q+\frac{4}{9}q^2\rt)$ \\[.5mm]
$3$ & $0$ & $\frac{1}{128}\lt(1+24q\rt)$			& $-\frac{135}{128}\lt(1+2q+\frac{4}{3}q^2+\frac{8}{27}q^3\rt)$ \\[.5mm]
$4$ & $0$ & $-\frac{5}{2048}\lt(1+48q+96q^2\rt)$		& $\frac{2835}{2048}\lt(1+\frac{8}{3}q+\frac{8}{3}q^3+\frac{32}{27}q^3+\frac{16}{81}q^4\rt)$\\[.5mm]
$1$ & $1$ & $-\4$				 		& $\4$ \\[.5mm]
$2$ & $1$ & $-\frac{1}{16}\lt(1-2q\rt)$				& $-\frac{3}{16}\lt(1+4q\rt)$ \\[.5mm]
$3$ & $1$ & $\frac{21}{128}\lt(1-\frac{12}{7}q-\frac{4}{7}q^2\rt)$ & $-\frac{45}{128}\lt(1-\frac{16}{3}q-4q^2\rt)$ \\[.5mm]
$2$ & $2$ & $-\frac{1}{16}$				 	& $\8$ \\[.5mm]
\hline
\end{tabular}
\caption{The non-vanishing coefficients of the series solutions.}
\label{sertab}
\end{center}
\end{table}
\begin{table}[h]
\begin{center}
\begin{tabular}{|c|c|l|l|}
\hline
$i$ &$ j$ & $c_{ij}$ & $d_{ij}$\\[.5mm]
\hline
$0$ & $0$ & $1$ 					& $1$ \\[.5mm]
$1$ & $0$ & $-\4$ 					& $\frac{9}{4}\lt(1-\frac{8}{9}q\rt)$ \\[.5mm]
$2$ & $0$ & $-\frac{1}{32}$				& $-\frac{25}{32}\lt(1-\frac{8}{25}q-\frac{8}{25}q^2\rt)$ \\[.5mm]
$3$ & $0$ & $\frac{5}{384}\lt(1+\frac{48}{5}q\rt)$	& $\frac{179}{384}\lt(1+\frac{288}{179}q-\frac{360}{179}q^2-\frac{32}{179}q^3\rt)$ \\[.5mm]
$4$ & $0$ & $-\frac{31}{6144}\lt(1+\frac{768}{31}q+\frac{816}{31}q^2\rt)$	& $-\frac{2095}{6144}\lt(1+\frac{1584}{419}q-\frac{1008}{419}q^2-\frac{512}{419}q^3-\frac{48}{419}q^4\rt)$\\[.5mm]
$1$ & $1$ & $\frac{1}{40}$				& $\frac{2}{15}$ \\[.5mm]
$2$ & $1$ & $\frac{37}{160}\lt(1+\frac{18}{37}q\rt)$	& $-\frac{87}{20}\lt(1+\frac{214}{261}q\rt)$ \\[.5mm]
$3$ & $1$ & $-\frac{193}{1280}\lt(1+\frac{564}{193}q+\frac{128}{193}q^2\rt)$ & $\frac{141}{20}\lt(1+\frac{117}{94}q+\frac{289}{564}q^2\rt)$ \\[.5mm]
$2$ & $2$ & $-\frac{29}{160}$				 & $\frac{23}{15}$\\[.5mm]
\hline
\end{tabular}
\caption{The non-vanishing coefficients of the logarithmic solutions.}
\label{logtab}
\end{center}
\end{table}
%
}             
%
%
%
%
\end{appendix}

\end{document}